\documentclass[twocolumn,aps,amssymb,superscriptaddress,showpacs,prb]{revtex4}
\usepackage{graphicx}
\usepackage{scalefnt}

\begin{document}

\title{Polar phonons and spin-phonon coupling in HgCr$_{2}$S$_{4}$ and CdCr$_2$S$_4$}

\author{T.~Rudolf}
\author{Ch.~Kant}
\author{F.~Mayr}
\author{J.~Hemberger}
\affiliation{Experimental Physics V, Center for Electronic Correlations and
Magnetism, University of Augsburg, D-86135~Augsburg, Germany}

\author{V.~Tsurkan}
\affiliation{Experimental Physics V, Center for Electronic Correlations and
Magnetism, University of Augsburg, D-86135~Augsburg, Germany}
\affiliation{Institute of Applied Physics, Academy of Sciences of Moldova,
MD-2028~Chi\c{s}in\u{a}u, Republic of Moldova}

\author{A.~Loidl}
\affiliation{Experimental Physics V, Center for Electronic Correlations and
Magnetism, University of Augsburg, D-86135~Augsburg, Germany}

\date{\today}

\begin{abstract}
Polar phonons of HgCr$_2$S$_4$ and CdCr$_2$S$_4$ are studied by far-infrared
spectroscopy as a function of temperature and external magnetic field.
Eigenfrequencies, damping constants, effective plasma frequencies and
Lyddane-Sachs-Teller relations, and effective charges are determined.
Ferromagnetic CdCr$_2$S$_4$ and antiferromagnetic HgCr$_2$S$_4$ behave rather
similar. Both compounds are dominated by ferromagnetic exchange and although
HgCr$_2$S$_4$ is an antiferromagnet, no phonon splitting can be observed at the
magnetic phase transition. Temperature and magnetic field dependence of the
eigenfrequencies show no anomalies indicating displacive polar soft mode
behavior. However, significant effects are detected in the temperature
dependence of the plasma frequencies indicating changes in the nature of the
bonds and significant charge transfer. In HgCr$_2$S$_4$ we provide experimental
evidence that the magnetic field dependence of specific polar modes reveal
shifts exactly correlated with the magnetization showing significant
magneto-dielectric effects even at infrared frequencies.
\end{abstract}

%\doi{}


\maketitle

\section{Introduction}

In the chromium spinels $A$Cr$_2X_4$ ($A$ = Zn, Cd, Hg; $X$ = O, S, Se)
competing antiferromagnetic (AFM) and ferromagnetic (FM)
interactions~\cite{baltzer} establish a fascinating phase diagram with complex
ground states.~\cite{rudolfNJP07} In these compounds the Cr$^{3+}$ ions exhibit
a half-filled $t_{2g}$ shell constituting a spin-only state with $S = 3/2$ and
vanishing spin-orbit coupling.

The oxides, with small lattice constants, are dominated by direct AFM Cr-Cr
exchange. As the chromium ions form a corner-sharing tetrahedral network, which
is a prototypical example of a pyrochlore lattice, these compounds are
geometrically frustrated~\cite{ramirez} and undergo antiferromagnetic spin
order at temperatures $T \ll \Theta_{CW}$, where $\Theta_{CW}$ is the
Curie-Weiss temperature. The magnetic transitions are accompanied by structural
distortions which have been explained in terms of a spin-driven Jahn-Teller
(JT) effect.~\cite{yamashita,tchernyshyov} In most of these AFM spinels, the
phonon eigenmodes reveal significant splittings below the AFM ordering
temperature.~\cite{rudolfNJP07,sushkov,hemberger2006,rudolfPRB07} The phonon
splittings are driven by magnetic exchange interactions via strong spin-phonon
coupling, despite the fact that Cr$^{3+}$ in an octahedral crystalline electric
field reveals a spherical charge distribution with a $g$-value close to $g =
2$, synonymous with the absence of conventional spin-orbit coupling.

At large lattice constants, in some of the sulfides and
selenides,~\cite{baltzer,rudolfNJP07} 90$^\circ$ Cr-$X$-Cr exchange becomes
dominant, leading to FM semiconductors with relatively high magnetic transition
temperatures. Recently relaxor ferroelectricity has been reported for
CdCr$_2$S$_4$ (Ref.~\onlinecite{hemberger2005}) and HgCr$_2$S$_4$.~\cite{weber}
This seems to be a further experimental evidence concerning the long-standing
question of local polar distortions in spinel compounds.~\cite{grimes} However,
the origin of these polar distortions is unclear~\cite{schmid} and in ab initio
phonon calculations~\cite{fennie} no indications for a structural instability
due to soft phonon modes have been detected. Recently, evidence accumulates
that the large magneto-capacitive (MC) effects in the chromium spinels strongly
depend on thermal treatment and/or doping. In pure ceramics, for example, MC
effects are absent or strongly reduced, but they are recovered in indium doped
samples. On the contrary, annealing in vacuum eliminates MC effects in Cl-doped
single crystals, but enables them in Cl-free single
crystals.~\cite{hemberger1065,tsurkanxx} The observation that marginal doping
with charge carriers induces dipolar relaxation phenomena reminds of similar
effects in transition-metal monoxides doped on a subpercent level with Li
(Ref.~\onlinecite{bosman}) or on metal deficient semiconducting
NiO.~\cite{lunkenheimer} Specifically for MnO it has been suggested that Li
doping leads to Mn$^{4+}$ ions with a strongly Jahn-Teller active $d^4$
electron configuration, inducing polaron-like bound charge carriers. In a
similar way, CdCr$_2$S$_4$ with sulphur deficiency or doped with Cl$^-$ will
exhibit Cr$^{2+}$ with a JT active $d^4$ configuration and possible dipolar
relaxation of bound polarons. Additional experiments will be necessary to
unravel if the observed relaxor phenomena can be explained by doping only or by
intrinsic lattice instabilities driven by off-centering of the Cr ions as has
been proposed by recent Raman scattering experiments.~\cite{lemmens}

It is known since long time that in CdCr$_2$S$_4$ strong spin-phonon coupling
is active,~\cite{wakamura88} leading to an unusual temperature dependence of
eigenfrequencies and dampings of the infrared (IR) active phonon modes. In
addition, recent Raman scattering experiments provide experimental evidence for
polar distortions well above the magnetic ordering temperature.~\cite{lemmens}
In this work we communicate the results of a detailed IR study. We report on
the temperature dependence of eigenfrequencies, dampings and effective plasma
frequencies of all IR active phonons in HgCd$_2$S$_4$ and CdCr$_2$S$_4$. Our
results show, that while there are indeed no indications of strong phonon
softening, we observe strong effects in the temperature dependence of the
plasma frequencies of specific modes, indicative for significant changes of the
character of bonds and that the effective charges undergo large variations as a
function of temperature. In addition it seems worthwhile to search for phonon
splitting in AFM HgCr$_2$S$_4$, as has been observed in the related Cr spinels
ZnCr$_2$S$_4$ (Ref.~\onlinecite{hemberger2006}) and ZnCr$_2$Se$_4$
(Ref.~\onlinecite{rudolfPRB07}) below the antiferromagnetic phase-transition
temperature. HgCr$_2$S$_4$ is close to FM order and moderate magnetic fields
well below 1~T induce a polarized ferromagnetic state which is shifted by
almost 60~K in external fields of 5~T.~\cite{tsurkan06} It thus seems
particularly important to study the spin-phonon coupling also as function of an
external magnetic field.

\section{Experimental details}

CdCr$_2$S$_4$ and HgCr$_2$S$_4$ crystallize in the normal cubic spinel
structure ($Fd\bar{3}m$) with lattice constants $a = 1.0247$~nm and $a =
1.0256$~nm and sulphur fractional coordinates $x = 0.263$ and $x = 0.267$ for
the cadmium and mercury compound, respectively.~\cite{rudolfNJP07} Both
compounds are dominated by strong FM exchange with Curie-Weiss temperatures of
the order of 150~K. But while CdCr$_2$S$_4$ reveals FM order at $T_C =
84.5$~K,~\cite{rudolfNJP07,hemberger2005} HgCr$_2$S$_4$ exhibits spiral-like
AFM spin order below 22~K.~\cite{rudolfNJP07,tsurkan06,chapon}

The IR experiments on CdCr$_2$S$_4$ were performed on as-grown single crystals
obtained by chemical transport reaction method using Cl as a transport agent.
The measurements were repeated on high-purity ceramic samples to check for
possible effects on the phonon properties by doping with chlorine. In the
single crystals, a Cl concentration $<0.25$\,mol\% and a concomitant charge
carrier density may be present, but from X-ray studies with synchrotron
radiation we certainly can exclude any clustering of Cl ions. From detailed
wave length sensitive microprobe analysis we are sure that single crystals and
ceramics investigated in these studies are very close to ideal stoichiometry.
Our IR results in CdCr$_2$S$_4$ are in good agreement with those reported by
Wakamura and Arai~\cite{wakamura88} on ceramic samples. CdCr$_2$S$_4$ reveals
relaxor-type ferroelectric behavior below approximately 130~K. This glass-like
transition is dynamic and strongly depends on the measuring
frequency.~\cite{hemberger2005} Soft ferromagnetism with a fully developed
magnetic moment of the chromium spins appears below 84.5~K. The measurements on
HgCr$_2$S$_4$, documented in this work, were made on ceramic pellets with
polished surfaces. HgCr$_2$S$_4$ is dominated by strong ferromagnetic
fluctuations,~\cite{tsurkan06} but undergoes AFM spin order below
22~K.~\cite{weber,tsurkan06,chapon} Ferroelectric hysteresis loops have been
reported close to 70~K, well above the AFM phase transition.~\cite{weber}

The reflectivity measurements were carried out in the far-infrared range using
the Bruker Fourier-transform spectrometer IFS 113v equipped with a He-bath
cryostat and with a split-coil magnet for measurements in external magnetic
fields up to 7~T. With this setup we were able to measure the frequency range
from 50 to 700~cm$^{-1}$ with high precision. The reflectivity measurements
were performed on polished surfaces of the samples. To get an independent
estimate of the background dielectric constant, the static dielectric constant
$\epsilon_0$ has been measured for both compounds between 150--500~GHz using
millimeter spectroscopy.~\cite{gorshunov} In these measurements the real and
imaginary part of the dielectric constant can be measured directly via
transmission and phase shift, without relying on any Kramers-Kronig-type
analysis for the calculation of the complex permittivity.

Room temperature far-infrared spectra of HgCr$_2$S$_4$ and CdCr$_2$S$_4$ were
published by Lutz and coworkers~\cite{lutz} and by Wakamura {\it et
al.}~\cite{wakamura80,wakamura88} Incomplete spectra of CdCr$_2$S$_4$ have also
been published by Lee.~\cite{lee} Some preliminary far-infrared results on
HgCr$_2$S$_4$ have been published in Ref.~\onlinecite{rudolfNJP07}. The
temperature dependence of the phonon eigenfrequencies and dampings in
CdCr$_2$S$_4$ has been reported by Wakamura and Arai.~\cite{wakamura88} The
present experiments were undertaken with special attention to search for
fingerprints in the phonon spectra for a possible onset of ferroelectricity in
both compounds and to search for phonon splittings in AFM HgCr$_2$S$_4$, which
could be expected on symmetry arguments alone. In addition, we also performed
far-infrared measurements as function of external magnetic fields to search for
magneto-dielectric effects at far-infrared frequencies.

\section{Analysis of the results}

\subsection{Modelling of the far-infrared spectra}

The phonon contribution of the complex dielectric function
$\epsilon(\omega)=\epsilon_1(\omega)+i\epsilon_2(\omega)$ of an insulating
crystal is obtained by calculating the factorized function
\begin{equation}
\label{equ1}
\epsilon(\omega)=\epsilon_{\infty}\prod_j\frac{\omega_{Lj}^2-\omega^2-i\gamma_{Lj}\omega}{\omega_{Tj}^2-\omega^2-i\gamma_{Tj}\omega}.
\end{equation}
Here $\omega_{Lj}$, $\omega_{Tj}$, $\gamma_{Lj}$ and $\gamma_{Tj}$
correspond to longitudinal (L) and transversal (T) eigenfrequencies
($\omega_j$) and dampings ($\gamma_j$) of mode $j$, respectively.
$\epsilon_{\infty}$ results from high-frequency electronic
absorption processes and can be experimentally determined from the
reflectivity or the index of refraction at frequencies larger than
the phonon eigenfrequencies. At normal incidence $\epsilon(\omega)$
is related to the reflectivity $R(\omega)$ via
\begin{equation}
\label{equ2}
R(\omega)=\left|\frac{\sqrt{\epsilon(\omega)}-1}{\sqrt{\epsilon(\omega)}+1}\right|^2.
\end{equation}
Using Eqs.~(\ref{equ1}) and (\ref{equ2}) the reflectivity spectra can be
unambiguously described using a four-parameter fit routine, which has been
recently developed by A. Kuzmenko.~\cite{kuzmenko} In most cases, when the
modes are well separated, this fit allows a precise determination of
$\epsilon_{\infty}$, longitudinal and transversal eigenfrequencies
($\omega_{Lj}, \omega_{Tj}$) and dampings ($\gamma_{Lj}, \gamma_{Tj}$). Using
this formalism, the dielectric strength $\Delta\epsilon$ can then be calculated
via
\begin{equation}
\label{equ3}
\Delta\epsilon=\epsilon_0-\epsilon_{\infty}=\sum_j\Delta\epsilon_j=\epsilon_{\infty}\left(\prod_j\frac{\omega_{Lj}^2}{\omega_{Tj}^2}-1\right),
\end{equation}
where $\epsilon_0$ corresponds to the {\it static}
$\epsilon(\omega\rightarrow0)$ dielectric constant and
$\Delta\epsilon_j$ to the strength of mode $j$, which in the case of
non-overlapping modes can be explicitely derived:
\begin{equation}
\label{equ4}
\Delta\epsilon_j=\epsilon_{\infty}\frac{\omega_{Lj}^2-\omega_{Tj}^2}{\omega_{Tj}^2}\prod_{i=j+1}\frac{\omega_{Li}^2}{\omega_{Ti}^2}.
\end{equation}
Assuming that the damping of transversal and longitudinal optical phonons is
identical, Eq.~(\ref{equ1}) reduces to
\begin{equation}
\label{equ5}
\epsilon(\omega)=\epsilon_{\infty}+\sum_j\frac{\omega_j^2\cdot\Delta\epsilon}{\omega^2_j-\omega^2-i\gamma_j\omega}.
\end{equation}
This is the well known Lorentzian lineshape with three independent parameters.
$\omega_j$ equals $\omega_{Tj}$ of the four-parameter model and
$\epsilon_{\infty}$ is an additional variable, which, of course, has to be
considered in both models. The mode strength $\Delta\epsilon$ can be related to
an effective ``ionic'' plasma frequency $\Omega$ with
\begin{equation}
\label{equ6} \Delta\epsilon_j\cdot\omega_{Tj}^2=\Omega_j^2,
\end{equation}
The effective plasma frequency of an IR active mode $j$, $\Omega_j$ can then be
related to the eigenfrequencies via
\begin{equation}
\label{equ7}
\Omega_j^2=(\omega_{Lj}^2-\omega_{Tj}^2)\cdot\tilde{\epsilon}_{\infty}(j)
\end{equation}
where $\tilde{\epsilon}_{\infty}$ is the appropriate high-frequency dielectric
constant for mode $j$, including all ionic contributions of modes $i$ with
$i>j$ (see Eq.~(\ref{equ4})):
\begin{equation}
\label{equ8}
\tilde{\epsilon}_{\infty}(j)=\epsilon_{\infty}\prod_{i=j+1}\frac{\omega_{Li}^2}{\omega_{Ti}^2}
\end{equation}
Finally, the dielectric strength of all polar modes can be related to a sum
over all effective plasma frequencies $\Omega_j$ with
\begin{equation}
\label{equ9}
\Omega^2=\sum_j\Omega_j^2=\sum_j\Delta\epsilon_j\omega_{Tj}^2
\end{equation}
In favorable cases, using Eqs.~(\ref{equ5})--(\ref{equ7}), the ionicity or
covalency of bonds involved in polar modes can be calculated. This is possible
on the basis of Eq.~(\ref{equ6}) if the pattern of displacements of a specific
phonon mode is known from an eigenvector analysis. But even if this is not the
case, one can arrive at definit conclusions about the effective charge of an
ion. It has been pointed out by Scott~\cite{scott} and has later been applied
by Wakamura~\textit{et al.}~\cite{wakamura80,wakamura88,wakamura81} to spinel
compounds that for a multi mode system one can assume
\begin{equation}
\label{equ10}
\Omega^2=\sum_j(\omega_{Lj}^2-\omega_{Tj}^2)\cdot\tilde\epsilon_{\infty}(j)=\frac{\epsilon_{\infty}}{V\epsilon_{vac}}\sum_k\frac{(Z_k^*e)^2}{m_k}
\end{equation}
Here $V$ is the unit-cell volume, $\epsilon_{vac}$ the dielectric permittivity
of free space, and $Z^*_ke$ is the effective charge of the $k$-th ion
contributing to a specific polar mode with mass $m_k$. The $k$ sum is over all
atoms in the unit cell. This equation has to be combined with the expression
for lattice charge neutrality
\begin{equation}
\label{equ11}
\sum_kZ^*_k=0.
\end{equation}
In ternary compounds Eqs.~(\ref{equ10}) and (\ref{equ11}) do not allow to
determine all charges of the ions unambiguously. In these cases specific
assumptions concerning the effective charges of at least one ion have to be
made.~\cite{wakamura80,wakamura88,wakamura81}

In the present investigation we are also interested in possible
soft-mode phenomena accompanying ferroelectric phase transitions.
Hence we will focus on the generalized Lyddane-Sachs-Teller (LST)
relation, which can be written as
\begin{equation}
\label{equ12}
\epsilon_0=\epsilon_{\infty}\prod_j\frac{\omega_{Lj}^2}{\omega_{Tj}^2}.
\end{equation}
In the case of a soft transverse optic mode the LST relation gives a diverging
static dielectric constant, which indeed is observed in many proper
ferroelectrics.

\subsection{Temperature dependence of eigenfrequencies and dampings}

\subsubsection{Anharmonic effects}

To get an estimate of the influence of the magnetic exchange interactions on
the phonon properties we tried to describe the purely anharmonic temperature
dependence of eigenfrequencies and dampings by a simple model, assuming
\begin{equation}
\label{equ13}
\omega_{Tj}=\omega_{T0j}\left(1-\frac{c_j}{\exp(\Theta/T)-1}\right)
\end{equation}
for the temperature dependence of the transverse eigenfrequencies and
\begin{equation}
\label{equ14}
\gamma_{Tj}=\gamma_{T0j}\left(1+\frac{d_j}{\exp(\Theta/T)-1}\right)
\end{equation}
for the temperature dependence of damping due to canonical anharmonic effects.

$\omega_{Tj}$ and $\gamma_{Tj}$ are eigenfrequency and damping of the
transversal optical mode $j$ and $\Theta$ is the Debye temperature, which has
been determined from an average of the four IR active phonon frequencies. For
the analysis of the anharmonic contributions, Debye temperatures of 361~K
(HgCr$_2$S$_4$) and 372~K (CdCr$_2$S$_4$) have been determined. $\omega_{T0j}$
and $\gamma_{T0j}$ are the values of phonon frequency and damping of the
transverse mode $j$ at 0~K. $c_j$ and $d_j$ are mere fitting parameters
determining the strength of the anharmonic contributions. Detailed calculations
of temperature and frequency dependent anharmonic contributions to
eigenfrequencies and dampings can be found in Ref.~\onlinecite{cowley}.

\begin{figure}
\includegraphics{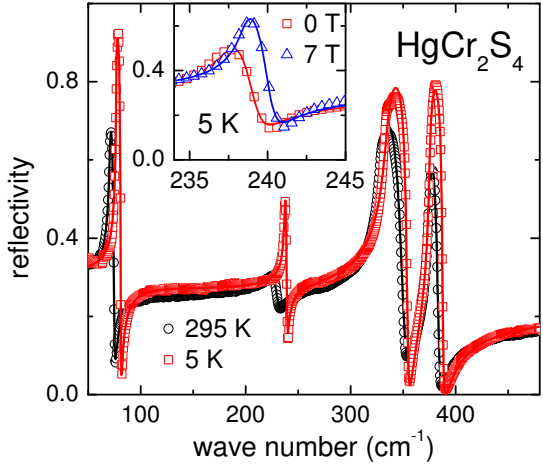}
\caption{\label{fig1}(Color online) Reflectivity vs wave number in
HgCr$_2$S$_4$ at 5~K (open red squares) and 295~K (open black circles). The
inset shows the phonon response close to 240~cm$^{-1}$ as measured in zero
magnetic field (red open squares) and in a magnetic field of 7~T (blue open
triangles). In external magnetic fields the excitation frequency is shifted
towards higher energies. The solid lines are the results of fits as described
in the text.}
\end{figure}

\subsubsection{Spin-phonon coupling}

The spin-phonon coupling in magnetic semiconductors has been described by
Baltensperger and Helman,~\cite{baltensperger1968}
Baltensperger,~\cite{baltensperger1970} Br\"{u}esch and D'Ambrogio,~\cite{brueesch}
Lockwood and Cottam~\cite{lockwood} and Wesselinowa and
Apostolov.~\cite{wesselinowa} These authors have shown that the frequency shift
of a given phonon mode as function of temperature is determined by a
spin-correlation function
\begin{equation}
\label{equ15}
\omega=\omega_0+\lambda <S_i \cdot S_j>.
\end{equation}
Here $\omega$ is the renormalized phonon frequency, $\omega_0$ denotes the
eigenfrequency in the absence of spin-phonon coupling and $\lambda$ the
spin-phonon coupling constant. On the basis of these theories Wakamura and
Arai~\cite{wakamura88} attempted to describe the experimentally observed
frequency shifts $\Delta\omega$ in CdCr$_2$S$_4$ by taking into account a sum
of two terms, namely due to FM and AFM exchange, respectively:
\begin{equation}
\label{equ16} \Delta\omega=\frac{-R_1<S_1 \cdot S_2> + R_2<S_1 \cdot
S_3>}{<S_0^z>^2}
% \Delta\omega=-R_1\frac{<S_1 \cdot S_2>}{<S_0^z>^2} + R_2\frac{<S_1 \cdot
% S_3>}{<S_0^z>^2}
\end{equation}
Here $R_1$ and $R_2$ are spin dependent force constants of the lattice
vibrations deduced as the squared derivatives of the exchange integrals with
respect to the phonon displacements.~\cite{wesselinowa} $R_1$ describes the
nearest-neighbor (nn) FM, $R_2$ the AFM next-nearest-neighbor (nnn) exchange.
Using this formalism, negative and positive frequency shifts are depending on
the strength of FM or AFM exchange interactions, respectively. In CdCr$_2$S$_4$
and HgCr$_2$S$_4$ we identify the negative contributions to the
eigenfrequencies with nn FM Cr-S-Cr exchange and positive contributions with
nnn AFM Cr-S-Cd/Hg-S-Cr exchange. Hence, vibrations involving Cr-S bonds will
predominantly be influenced by FM exchange, while eigenmodes of Cd-S or Hg-S
ions will be sensitive to AFM exchange.

The temperature dependence of the phonon damping due to spin-phonon
interactions has been calculated by Wesselinova and
Apostolov.~\cite{wesselinowa} These authors show that in the magnetically
ordered phase an additional contribution to the phonon damping arises due to
spin-phonon interactions, which vanishes in the paramagnetic phase.

\begin{figure}
\includegraphics{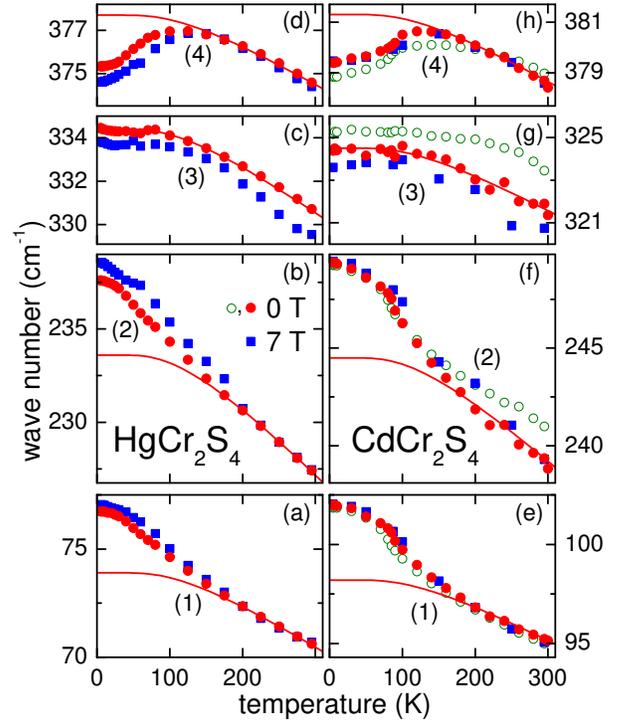}
\caption{\label{fig2}(Color online) Temperature dependencies of
eigenfrequencies (transverse optical modes) of HgCr$_2$S$_4$ (left frames
(a)--(d)) and CdCr$_2$S$_4$ (right frames (e)--(h)). The four phonon modes are
labeled with increasing wave number. Eigenfrequencies in zero external field
are shown as closed red circles, in magnetic fields of 7~T as closed blue
squares. Eigenfrequencies in CdCr$_2$S$_4$ as determined in zero magnetic field
in a ceramic sample are shown by open green circles. The red solid lines are
fits to the eigenfrequencies in zero field assuming a purely anharmonic
temperature dependence of the eigenfrequencies utilizing a fixed Debye
temperature (Eq.~(\ref{equ13}), see text).}
\end{figure}

\section{Experimental results}

\subsection{Reflectivity measurements in zero external field}

Figure~\ref{fig1} shows the reflectivity $R$ of HgCr$_2$S$_4$ as measured on
ceramic samples at 295~K (black circles) and at 5~K (red squares). Already this
figure provides clear evidence of a significant temperature dependence of the
eigenfrequencies and an enormously strong temperature dependence of damping and
dielectric strength of the modes. In a normal anharmonic solid one expects an
almost temperature independent dielectric strength, on increasing temperature
eigenfrequencies should moderately decrease, whereas damping increases. The
solid lines through the measured reflectivity correspond to the four-parameter
fit using Eqs.~(\ref{equ1}) and~(\ref{equ2}). This fit provides a reasonable
description of the experimental results. Deviations appear only at the maximum
of mode~3 close to 330~cm$^{-1}$, which probably arise due to imperfections of
the sample surface or due to multi-phonon processes.

\begin{figure}
\includegraphics{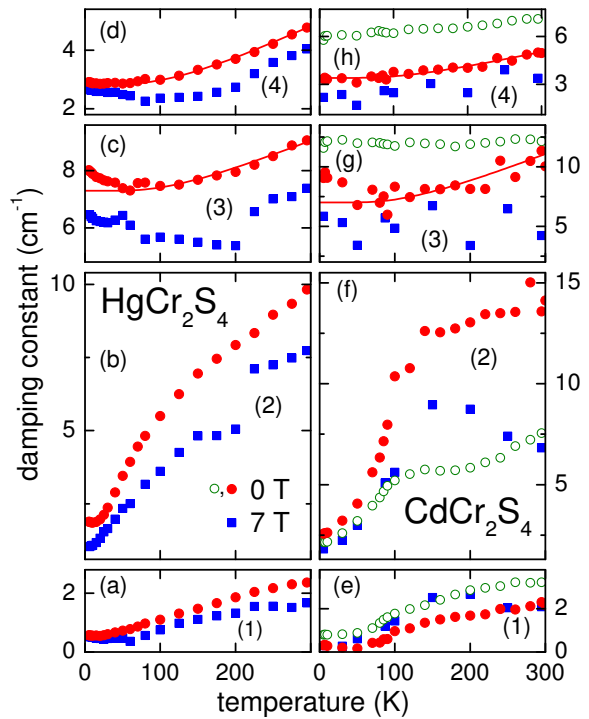}
\caption{\label{fig3}(Color online) Phonon damping of the transverse optical
modes versus temperature for HgCr$_2$S$_4$ (left frames (a)--(d)) and
CdCr$_2$S$_4$ (right frames (e)--(h)). Damping constants in zero external field
are shown as closed red circles, in magnetic fields of 7~T as closed blue
squares. Damping constants in CdCr$_2$S$_4$ as determined in a ceramic sample
are shown by open green circles. The solid lines are fits assuming a purely
anharmonic temperature dependence of the damping utilizing a fixed Debye
temperature (see text).}
\end{figure}

The main results of the analysis of the reflectivity spectra of ceramic
HgCr$_2$S$_4$ and single-crystalline CdCr$_2$S$_4$ as a function of temperature
are shown in Figs.~\ref{fig2} and~\ref{fig3}. Figure~\ref{fig2} shows the
eigenfrequencies for HgCr$_2$S$_4$ (Fig.~\ref{fig2}(a)--(d)) and CdCr$_2$S$_4$
(Fig.~\ref{fig2}(e)--(h)) and Fig.~\ref{fig3}(a)--(h) the corresponding damping
of all modes. The modes $j$ are labeled with the numbers~1 to~4 with increasing
wave number. Only parameters of the transverse optical modes are shown.

\begin{table}\caption{\label{tab1}Transversal $\omega_{T}$ [cm$^{-1}$] and longitudinal $\omega_{L}$ [cm$^{-1}$] optical eigenfrequencies of CdCr$_2$S$_4$ compared with theoretical values predicted by LSDA\,+\,U (Ref.~\onlinecite{fennie}). The extrapolated eigenfrequencies at 0~K $\omega_{T0j}$ [cm$^{-1}$] in the absence of spin-phonon coupling are also listed.}
\begin{ruledtabular}
\begin{tabular}{cccccc}
& \multicolumn{3}{c}{Experiment} & \multicolumn{2}{c}{LSDA\,+\,U}\\
mode & $\omega_{Tj}$ & $\omega_{Lj}$ & $\omega_{T0j}$ & $\omega_{Tj}$ & $\omega_{Lj}$\\
 \hline
(1) & 102 & 105 & 98 & 104 & 107\\
(2) & 249 & 251 & 244 & 249 & 251\\
(3) & 324 & 352 & 324 & 339 & 362\\
(4) & 379 & 395 & 381 & 385 & 398\\
\end{tabular}
\end{ruledtabular}
\end{table}

In the right frames of Figs.~\ref{fig2} and~\ref{fig3} we also included
eigenfrequencies and phonon damping for CdCr$_2$S$_4$ as obtained on
high-purity ceramic samples. Ceramic CdCr$_2$S$_4$ has been investigated to
determine a possible influence of the doping with Cl which can not be fully
avoided during the single crystal growth process. With respect to the
eigenfrequencies, the agreement between single crystalline and ceramic results
is reasonable. Only mode~3 shows slight deviations of the order of 1~cm$^{-1}$.
However, we have to keep in mind that the fit of mode~3 is not perfect (see
Fig.~\ref{fig1}) and the apparent differences probably can be attributed to the
fitting procedure. More significant deviations appear in the temperature
dependence of the damping constants. While the general temperature dependence
is rather similar, specifically mode~2 and~4 reveal rather different absolute
values. Nevertheless, we conclude from these results that while a small amount
of doping significantly influences the dielectric properties at low
frequencies, the phonon properties seen via the reflectivity are only barely
influenced. In what follows we only will discuss the results obtained on
ceramic HgCr$_2$S$_4$ and single-crystalline CdCr$_2$S$_4$.

The eigenfrequencies for HgCr$_2$S$_4$ (Fig.~\ref{fig2}(a)--(d)) span a range
from 70--377~cm$^{-1}$, for CdCr$_2$S$_4$ (Fig.~\ref{fig2}(e)--(h)) from 95 to
381~cm$^{-1}$. This observation indicates that the high-frequency modes in both
compounds exhibit almost identical eigenfrequencies and Cr and S ions can be
involved only in mode~4. The frequencies of phonon modes~1 scale with the
square root of the inverse masses of mercury and cadmium, pointing towards the
fact that in the low-frequency modes mainly the closed-shell metals of the zinc
group are involved. Striking similarities between antiferromagnetic
HgCr$_2$S$_4$ and ferromagnetic CdCr$_2$S$_4$ can also be recognized for the
temperature dependence of the damping constants. At low temperatures the
damping constants of both compounds are small for modes~1, 2~and~4, but rather
large for mode~3. Mode~2 reveals a strikingly strong temperature dependence for
HgCr$_2$S$_4$ as well as for CdCr$_2$S$_4$.

The solid lines in Figs.~\ref{fig2} and~\ref{fig3} correspond to the expected
anharmonic temperature dependence of frequencies and dampings as calculated
according to Eqs.~(\ref{equ13}) and~(\ref{equ14}). These curves have been
derived by fitting the high-temperature ($T > 150$~K) values of the
eigenfrequencies and dampings. This procedure seems to be straight forward and
correct for the eigenfrequencies of all modes in both compounds (Fig. 2). It
also properly describes the temperature dependence of the phonon dampings of
modes~3 and~4 for HgCr$_2$S$_4$ and CdCr$_2$S$_4$. The temperature dependence
of the damping constants of modes~1 and~2 of both compounds can not be
described in this way. For these two modes a continuous increase of the damping
is enhanced by a bump-like feature close to 150~K, which could result from
structural instabilities which appear definitely far above the magnetic
ordering temperatures.~\cite{hemberger2005,weber,lemmens} We would like to
recall that according to theory~\cite{wesselinowa} one would expect additional
contributions to the damping from phonon-magnon scattering in the magnetically
ordered phases, which vanish in the paramagnetic states. This definitely is not
observed. These strong cusplike features in the damping constants of both
compounds are rather extended and definitely do neither exhibit a maximum close
to $T_C$ in CdCr$_2$S$_4$ ($T_C = 84.5$~K) nor close to 60~K in HgCr$_2$S$_4$,
where in zero external field the strongest ferromagnetic fluctuations are
observed.~\cite{tsurkan06} Hence we think that an explanation of the damping
anomalies of mode~2 of both compounds in terms of ferromagnetic fluctuations is
rather unlikely. In recent Raman scattering experiments~\cite{lemmens}
pronounced anomalies in intensity and frequency of Raman-active modes have been
observed, arising at an onset temperature $T^*\approx130$~K, which coincides
with striking anomalies in the IR properties.

Already at first sight it becomes clear that AFM HgCr$_2$S$_4$ and FM
CdCr$_2$S$_4$ behave similar: The eigenfrequencies of modes~1 and~2 in both
compounds reveal clear positive shifts as compared to normal anharmonic
behavior, while modes~3 and~4 exhibit negative shifts when entering the
magnetic phase. This again documents that in modes~3 and~4 the eigenfrequencies
are determined mainly from force constants between Cr and S ions, which are
coupled ferromagnetically. On the other hand, Cd/Hg-S ions are involved in the
vibrations of modes~1 and~2, which obviously are strongly influenced by the AFM
Cr-S-Cd/Hg-S-Cr exchange. It is important to note that the temperature
dependence of the eigenfrequencies looks very similar for both compounds,
despite the fact that HgCr$_2$S$_4$ reveals antiferromagnetic order below 22~K,
while CdCr$_2$S$_4$ is ferromagnetic below 84.5~K. No indications of phonon
splittings in antiferromagnetic HgCr$_2$S$_4$ are observable. However, both
compounds show strong ferromagnetic fluctuations already at rather elevated
temperatures and it is probably this fact which explains the similarity in the
temperature dependence of the phonons. Strong anomalous contributions are also
observed in the temperature dependence of the damping. Modes~1 and 2 exhibit a
significant decrease of the damping below 150~K for both compounds. This
unusual behavior could be attributed to the onset of local polar order. On the
contrary, the dampings of modes~3 slightly increase below 100~K, while modes~4
of HgCr$_2$S$_4$ and CdCr$_2$S$_4$ are very close to the normal temperature
dependence of the damping in anharmonic solids.

\begin{figure}
\includegraphics{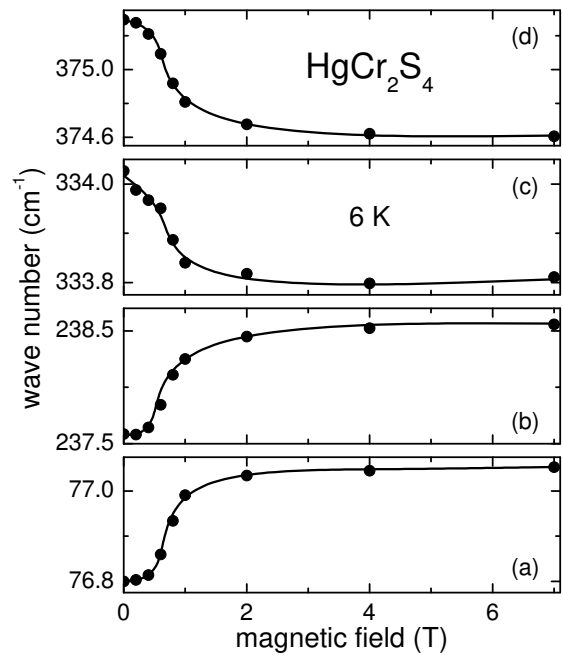}
\caption{\label{fig4}Magnetic field dependence of the eigenfrequencies of
HgCr$_2$S$_4$ for modes~1 (a) to~4 (d) at 6~K. The lines are drawn to guide the
eye.}
\end{figure}

The eigenfrequencies at 5~K are listed in Tab.~\ref{tab1} and compared to the
results from LSDA\,+\,U calculations of Fennie and Rabe.~\cite{fennie} Here we
also included extrapolated eigenfrequencies $\omega_{T0j}$ (see
Eq.~(\ref{equ13})), which correspond to the eigenfrequencies at 0~K in the
absence of spin-phonon coupling. With the exception of mode~3 we find good
agreement. From Tab.~\ref{tab1} we can also infer that the frequency shift at
0~K in CdCr$_2$S$_4$ due to spin-phonon coupling $\Delta\omega_{Tj} =
\omega_{Tj} - \omega_{T0j}$ is positive and of the order of 5~cm$^{-1}$ for
modes~1 and~2, but negative and significantly smaller for modes~3 and~4. The
very same is true for the spin-phonon coupling in HgCr$_2$S$_4$
(Fig.~\ref{fig2}(a)--(d)). This signals, as outlined above, that the modes~3
and~4 exhibiting negative shifts are only weakly influenced by FM 90$^{\circ}$
Cr-S-Cr exchange, while modes~1 and~2 are strongly influenced by AFM nnn
Cr-S-Cd/Hg-S-Cr exchange. This correlates nicely with calculations showing that
in the vibrational pattern of modes~3 and~4 only Cr and S ions are involved,
while a more complex pattern of motion, including all three types of ions,
characterizes modes~1 and~2.~\cite{fennie}

\subsection{Measurements in external magnetic fields}

On HgCr$_2$S$_4$ ceramics and on CdCr$_2$S$_4$ single crystals additional
reflectivity measurements have been performed in external magnetic fields up to
7~T. The inset of Fig.~\ref{fig1} shows mode~2 of HgCr$_2$S$_4$ at 5~K in zero
magnetic field compared to measurements in a field of 7~T. In an external field
of 7~T this mode is shifted to higher frequencies by an amount of approximately
1~cm$^{-1}$. The temperature dependencies of eigenfrequencies and damping
constants as obtained in both compounds for all modes at 7~T are included in
Figs.~\ref{fig2} and~\ref{fig3} (blue squares). As expected for a ferromagnet
with a high ordering temperature, the eigenfrequencies remain unchanged within
experimental uncertainties for CdCr$_2$S$_4$ (right frames of Fig.~\ref{fig2}).
For antiferromagnetic HgCr$_2$S$_4$, which shows a metamagnetic transition
below 1~T,~\cite{tsurkan06} shifts of the order of 1~cm$^{-1}$ are observed for
$T < 100$~K. The shifts are positive for modes~1 and~2 (see Fig.~\ref{fig2}(a)
and~(b)), but negative for modes~3 and~4 (see Fig.~\ref{fig2}(c) and~(d)). In
mode~3 (Fig.~\ref{fig2}(c)) the magnetic-field dependent shifts also show up at
room temperature. Again we have to remind that the fits of mode~3 have some
uncertainty due to a double-peak structure and in this case the differences of
the eigenfrequencies as a function of field may signal the experimental
uncertainties.

\begin{figure}
\includegraphics{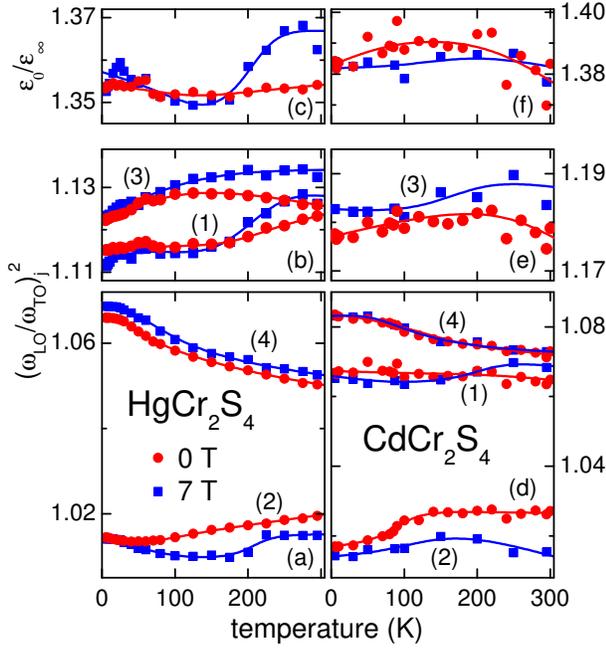}
\caption{\label{fig5}(Color online) Temperature dependence of the
Lyddane-Sachs-Teller relation for all modes (lower four frames) and the
generalized LST relation (upper two frames) for HgCr$_2$S$_4$ (left frames
(a)--(c)) and CdCr$_2$S$_4$ (right frames (d)--(f)) at zero magnetic fields
(closed red circles) and in an external magnetic field of 7~T (closed blue
squares). The lines are drawn to guide the eye.}
\end{figure}

The most precise measurements of the magnetic-field dependence of the phonon
eigenfrequencies can be made at a constant temperature and fixed spectrometer
setting. As representative example, Fig.~\ref{fig4} shows the field dependence
of all eigenmodes for HgCr$_2$S$_4$ at 6~K. As discussed above, the
eigenfrequencies of modes~1 and~2 (Fig.~\ref{fig4}(a),~(b)) increase, while the
frequencies of modes~3 and~4 (Fig.~\ref{fig4}(c),~(d)) decrease with increasing
magnetic field. The frequency shifts are of the order of 1~cm$^{-1}$ for
mode~2, but smaller for the other modes. The strongest gradients for all modes
appear well below 1~T, where AFM HgCr$_2$S$_4$ reveals a transition into a
polarized ferromagnetic state.~\cite{tsurkan06}

Much stronger effects as a function of magnetic field are observed for the
damping constants (see Fig.~\ref{fig3}). For all modes in both compounds we can
state that, while the overall temperature dependence remains similar with the
same characteristic trends, we observe rather large effects in the absolute
values. The reasion of these significant field dependencies of the phonon
damping is unclear at present and further investigations are necessary to
clarify this question.

\begin{figure}
\includegraphics{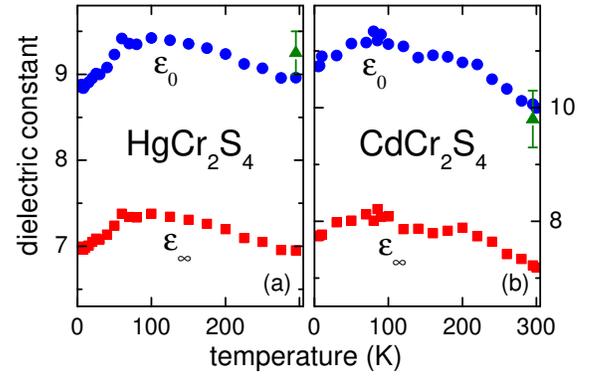}
\caption{\label{fig6}(Color online) Temperature dependencies of $\epsilon_0$
(closed blue circles) and $\epsilon_{\infty}$ (closed red squares) for
HgCr$_2$S$_4$ (left frame) and CdCr$_2$S$_4$ (right frame). Room-temperature
values of $\epsilon_{\infty}$ as obtained by THz spectroscopy are included
(closed green triangles).}
\end{figure}

\section{Analysis and Discussion}

With a more elaborated analysis we try to get insight into a possible polar
short-range-order phase transition above the magnetic phase transition
temperatures in both compounds. To do so we calculated effective plasma
frequencies and LST relations. The lower frames of Fig.~\ref{fig5} show the
Lyddane-Sachs-Teller relation for modes~1 to~4 for both compounds, the upper
two frames of Fig.~\ref{fig5} display the generalized LST relation for the
static dielectric constant as calculated via Eq.~(\ref{equ12}). Within 2\,\%
these relations remain constant for all modes. Hence, no significant softening
of TO modes becomes apparent, and it is obvious that a phonon softening of the
transverse acoustic modes, as observed in conventional ferroelectrics, plays no
role in HgCr$_2$S$_4$ and CdCr$_2$S$_4$. For the latter compound this has
already been concluded from first principle calculations.~\cite{fennie}
However, it seems worthwhile to note that beyond experimental uncertainties the
LST relation of mode~4 of both compounds shows a slight increase, signaling a
slight hardening of LO or softening of TO modes obviously related to
spin-phonon coupling. In Fig.~\ref{fig5} we also included the results of
measurements in an external field of 7~T. Within experimental uncertainty the
results agree with those obtained at zero field.

Figure~\ref{fig6} shows the temperature dependence of $\epsilon_0$ and
$\epsilon_{\infty}$ for HgCr$_2$S$_4$ (left frame) and CdCr$_2$S$_4$ (right
frame). Here we also included the results of $\epsilon_0$ obtained by THz
spectroscopy at room temperature. Both dielectric constants reveal only a
moderate temperature dependence. It is clear that the strong effects in the
temperature dependence of the dielectric constants as observed in
radio-frequency measurements~\cite{hemberger2005,weber} are not mirrored in
these far-infrared results. Those strong effects must be of relaxational origin
and obviously are related to slow polarization fluctuations. At room
temperature, $\epsilon_{\infty}$ is approximately 7.0 for HgCr$_2$S$_4$ and for
CdCr$_2$S$_4$. For HgCr$_2$S$_4$, $\epsilon_{\infty}$ has also been determined
by Lutz {\it et al.}~\cite{lutz} as 7.4, compared to 8.5 as obtained by
Wakamura.~\cite{wakamura88} For CdCr$_2$S$_4$, room temperature values of
$\epsilon_{\infty} = 6.9$~(Ref.~\onlinecite{lutz}),
7.6~(Ref.~\onlinecite{wakamura88}) and 7.84~(Ref.~\onlinecite{lee}) have been
reported. In insulating solids the dielectric constant at frequencies beyond
the ionic vibrations is determined by the electronic polarizability and the
band gap. The electronic polarizability is expected to be somewhat larger in
the mercury compound and despite minor effects due to thermal expansion
practically temperature independent. But $\epsilon_{\infty}$ also scales with
the inverse squared band gap. The band gap is of the order of 1.4~eV for
HgCr$_2$S$_4$ with a smooth red shift on decreasing temperature.~\cite{lehmann}
In CdCr$_2$S$_4$ at room temperature the band gap has been estimated as 1.57~eV
exhibiting an unusual blue shift on decreasing temperatures.~\cite{harbeke}
These band gaps could not explain the observed temperature dependence of
$\epsilon_{\infty}$ of the two compounds under consideration which are very
similar and suggest a red shift of the band edges from room temperature down to
100~K and a subsequent blue shift in the ferromagnetic phase (CdCr$_2$S$_4$) or
in the state dominated by ferromagnetic fluctuations (HgCr$_2$S$_4$). And
indeed, it has been suggested, that these reports on the band gaps in the
chromium spinels are heavily influenced by the Cr$^{3+}$ crystal field
excitations and do not represent the true electronic band
edges.~\cite{wittekoek}

\begin{figure}
\includegraphics{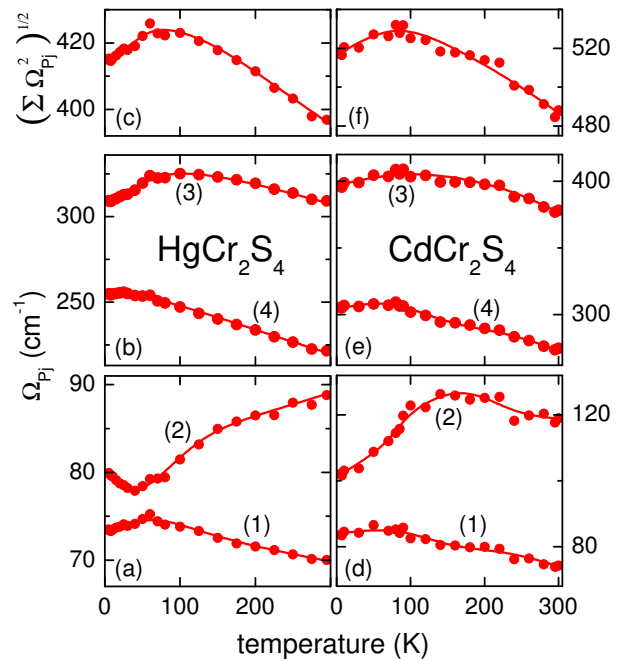}
\caption{\label{fig7}(Color online) Temperature dependence of the effective
plasma frequencies for modes~1 to~4 (lower four frames) and for the sum over
all plasma frequencies (upper two frames) of HgCr$_2$S$_4$ (left frames) and
CdCr$_2$S$_4$ (right frames). The lines are drawn to guide the eye.}
\end{figure}

In a second step, using Eqs.~(\ref{equ1}) and (\ref{equ7}) we calculated the
effective ``ionic'' plasma frequencies which, aside from thermal expansion
corrections, only depend on the effective charges of the ions under
consideration. The lower four frames of Figure~\ref{fig7} show the temperature
dependence of the plasma frequencies for HgCr$_2$S$_4$ and CdCr$_2$S$_4$ for
all modes, the upper two frames indicate the sum over all plasma frequencies in
zero external magnetic field. The plasma frequencies of both compounds for all
modes except for mode~2 show a moderate increase on decreasing temperature,
some saturate or go through a slight maximum close to 100~K. These effects are
of the order of 10\,\%. The temperature dependence of mode~2 of both compounds
is significantly different and strongly decreases below 150~K. In CdCr$_2$S$_4$
the overall effect of this mode is of the order of 25\,\% indicating strongly
decreasing effective charges towards low temperatures. This can only be
explained assuming strong charge transfer and/or that the character of the
bonds changes considerably. It could be a first experimental hint that charge
ordering phenomena are responsible for the appearance of local polar order.
Local charge order is thought to be responsible for the local polar state in
La:SrMnO$_3$ (Ref.~\onlinecite{mamin}) and charge-stripe order is the origin of
electronic ferroelectricity in LuFe$_2$O$_4$.~\cite{zhang} The sum over all
effective ionic plasma frequencies is dominated by the strongest modes~3 and~4
and reveals the over all behavior with a shallow maximum close to 100~K. The
larger ($\approx 25$\,\%) plasma frequency of CdCr$_2$S$_4$ as compared to the
mercury compound indicates a stronger ionicity for the former compound, a fact
that also follows from general considerations. According to Pauling's values of
electronegativity, both Cr-S and Hg-S form rather mixed bonds, however, with a
significant increase of covalency for the latter.

\begin{figure}
\includegraphics{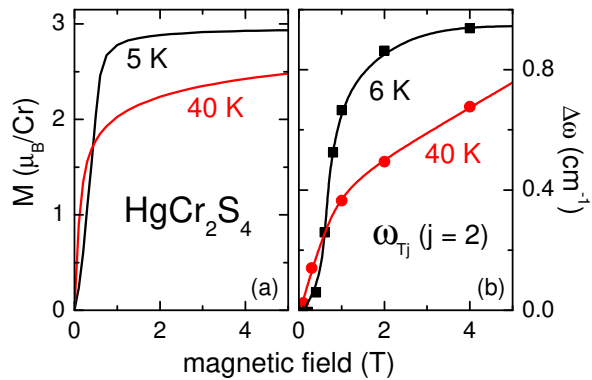}
\caption{\label{fig8}(Color online) Left frame (a): magnetic field dependence
of the magnetization in HgCr$_2$S$_4$ at 5~K and at 40~K, which is well below
and above the antiferromagnetic ordering temperature at 22~K. Right frame (b):
magnetic field dependence of the frequency shift of the transverse optical
phonon mode~2 at 6~K and 40~K. The lines are drawn to guide the eye.}
\end{figure}

The plasma frequencies of CdCr$_2$S$_4$, as derived from these experiments at
5~K, can be compared to those calculated by Fennie and Rabe~\cite{fennie} in
their LSDA\,+\,U approach. Good agreement is found for modes~1 - 3, however for
mode~4 at low temperatures we experimentally determined $\Omega =
309$~cm$^{-1}$, while theoretically 202~cm$^{-1}$ has been calculated. In
addition, from our experiments we determined $\Delta\epsilon$
(Eq.~(\ref{equ3})) of the order of 2.9 at 5~K compared to the theoretical value
of 2.3. Our value is in good agreement with the result by Wakamura \textit{et
al.},~\cite{wakamura80} who found $\Delta\epsilon = 2.8$. Corresponding to the
smaller plasma frequencies and to the higher covalency, $\Delta\epsilon$ for
HgCr$_2$S$_4$ is significantly smaller when compared to CdCr$_2$S$_4$ and is of
the order of~2 for all temperatures (see Fig.~\ref{fig6}(a)).

With the experimentally determined effective plasma frequencies as documented
in Fig.~\ref{fig7}, we tried to get some insight into the bonding of
HgCr$_2$S$_4$ and CdCr$_2$S$_4$ using Eqs.~(\ref{equ10}) and~(\ref{equ11}).
Utilizing Eq.~(\ref{equ10}), we can directly calculate the effective plasma
frequencies of both compounds assuming the nominal valencies $Z$ with
Hg$^{2+}$/Cd$^{2+}$, Cr$^{3+}$ and S$^{2-}$. For 300~K and with an averaged
$\epsilon_{\infty}$ we calculated theoretical ionic plasma frequencies of
1568~cm$^{-1}$ for HgCr$_2$S$_4$ and of 1517~cm$^{-1}$ for CdCr$_2$S$_4$. These
values have to be compared with the results shown in Figs.~\ref{fig7}(c)
and~\ref{fig7}(f), where we find an average value of the effective plasma
frequency of 415~cm$^{-1}$ for the mercury and of 515~cm$^{-1}$ for the cadmium
compound. Hence, we can state that the overall ionicity, i.e. the normalized
effective charge, $Z^*/Z$ amounts 26\,\% in HgCr$_2$S$_4$ and 34\,\% in
CdCr$_2$S$_4$. Proceeding one step further one may calculate the effective
charges of the different ions in both compounds combining Eqs.~(\ref{equ10})
and~(\ref{equ11}). However, due to the fact that three unknown quantities,
namely $Z^*_k$ for each ion, correspond to two equations only, we have to make
another ad hoc assumption, namely that the electronegativity of the closed
shell transition metal equals that of Cr. In other words, the relation of the
effective charges has to be $Z^*_{\scalefont{0.7}{\textrm{Hg/Cd}}} = 2/3\,
Z^*_{\scalefont{0.7}{\textrm{Cr}}}$, or equivalently
$Z^*_{\scalefont{0.7}{\textrm{Hg/Cd}}} = -Z^*_{\scalefont{0.7}{\textrm{S}}}$.
This produces an average reduction of the effective charges of all ions
according to the values determined from the overall theoretical and
experimental plasma frequencies. We achieve the effective charges for (Hg/Cd,
Cr, S) as (0.52, 0.78, -0.52) for the mercury and (0.66, 0.99, -0.66) for the
cadmium compound.

The effective plasma frequencies as obtained in an external magnetic field of
7~T are not shown. The magnetic field dependence of the far-infrared
intensities does not scale significantly enough with the external magnetic
field to compare results on different runs with different adjustments of the
spectrometer. However, as outlined above, the sensitivity is good enough to
search for changes of eigenfrequencies, dampings and intensities at a given
temperature as a function of external magnetic field without changing the
instrument setting. A representative example of such a measurement is given in
the inset of Fig.~\ref{fig1}. With a given setting and at constant temperature
we measured the magnetic field dependence of phonon mode~2 in HgCr$_2$S$_4$ at
6~K in the AFM state ($T<T_N$) and at 40~K in the paramagnetic state ($T>T_N$).
The results are shown in the right frame of Fig.~\ref{fig8} and compared to the
magnetization at similar temperatures (Fig.~\ref{fig8}(a)). The frequency shift
of the phonon mode is strongly correlated with the magnetization, even
revealing a strong meta-magnetic transition in the AFM state. Thus, mutual
correspondence of magnetization and phonon shift documents a significant
magneto-electric coupling even at optical frequencies.

\section{Concluding remarks}

Our far-infrared study of HgCr$_2$S$_4$ and CdCr$_2$S$_4$ was in part motivated
to search for experimental evidence for a polar phase transition close to
100~K, above the onset of magnetic order. The onset of relaxor ferroelectricity
has been brought up in Refs.~\onlinecite{hemberger2005} and~\onlinecite{weber}
and has reinforced old speculations of polar distortions in spinel compounds at
elevated temperatures.~\cite{grimes,schmid} For CdCr$_2$S$_4$ negative thermal
expansion below 150~K,~\cite{goebel,krimmel} the broadening of Bragg
reflections in X-ray diffraction experiments at temperatures above the
ferromagnetic phase transition at $T_C$ (Ref.~\onlinecite{goebel}) provided
some experimental evidence for short-range structural distortions close to
150~K. In recent synchrotron radiation experiments a (200) reflection, which is
forbidden in the spinel structure, showed up below 200~K with increasing
intensity towards 0~K.~\cite{krimmel} The onset of polar distortions also has
been claimed in recent Raman scattering experiments,~\cite{lemmens} where
pronounced anomalies in the Raman modes appear below a characteristic
temperature $T^*\approx130$~K.

We certainly can state that in accord with recent LSDA\,+\,U
calculations~\cite{fennie} we found no evidence for soft-mode behavior, which
usually is a fingerprint for ferroelectricity in proper ferroelectrics. This
statement is based on the observations as presented in Fig.~\ref{fig2}, but
also from the almost constant LST relations as presented in
Figs.~\ref{fig5}((c), (f)). However, we would like to recall that the soft-mode
concept, which couples a soft transverse mode to a divergent static dielectric
susceptibility in multiferroics may not be the adequate description. For the
perovskite-type family of mutliferroics it has been shown
theoretically~\cite{katsura,chupis} and proven
experimentally~\cite{pimenov,pimenov06} that electromagnons, i.e. coupled
spin-phonon excitations, are the relevant excitations. The ferroelectric
transition in these compounds is then rather driven by a soft magnetic
excitation.

Nevertheless, in this work we provide evidence for strong spin-phonon coupling.
The onset of these effects (Figs.~\ref{fig2} and~\ref{fig3}) certainly can be
detected far above the magnetic transition temperature and it is unlikely that
these effects are driven by ferromagnetic fluctuations. In both compounds
mode~2 behaves rather unusual, with a broad cusp in the temperature dependence
of the damping constant and specifically a significant anomaly in the plasma
frequency. Both anomalies are located close to 150~K.

In summary, the main outcome of this work is, that\\
i) within the experimental resolution, there is no splitting of phonon modes in
AFM HgCr$_2$S$_4$, and it behaves very similar like FM CdCr$_2$S$_4$,
demonstrating that both compounds are dominated by ferromagnetic exchange
interactions.\\
ii) There are strong frequency shifts in the phonon modes from normal
anharmonic behavior, which are positive for modes~1 and~2, but negative for
modes~3 and~4. For both compounds these shifts appear well above the onset of
magnetic order. For CdCr$_2$S$_4$, a detailed phonon study in zero magnetic
field on ceramic samples has been reported already by Wakamura and Arai some
decades ago.~\cite{wakamura88} The results of both studies are in good
agreement, however, with differences in the temperature dependence of the
dielectric
constants $\epsilon_0$ and $\epsilon_{\infty}$.\\
iii) In AFM HgCr$_2$S$_4$ we
find significant shifts of the eigenfrequencies by external magnetic fields,
which scale perfectly well with the macroscopic
magnetization.\\
iv) The plasma frequencies of both compounds reveal a moderate increase on
decreasing temperature with the exception of mode~2, which exhibits a strong
decrease below 150~K.\\
v) The Lyddane-Sach-Teller relations of all modes behave rather normal as
function of temperature and magnetic field. Within experimental uncertainty,
only mode~4 reveals a moderate increase in the LST relation in both
compounds.\\
vi) The overall ionicity as determined from the sum over all ionic effective
plasma frequencies is 28\,\% for HgCr$_2$S$_4$ and 33\,\% for CdCr$_2$S$_4$,
under the assumption that the normal valency would produce 100\,\% ionic bonds.

\begin{acknowledgements}
Stimulating discussions and helpful comments by P. Lemmens are gratefully
acknowledged. This work partly was supported by the Deutsche
Forschungsgemeinschaft through the German Research Collaboration SFB~484
(University of Augsburg).
\end{acknowledgements}

\end{document}